\documentclass[
   ,final            
  ]
  {aipproc}

\layoutstyle{8x11single}

\begin{document}

\title{The Qweak Experiment -- A Search For New Physics At The TeV Scale By
Measurement Of The Proton's Weak Charge}



\author{W.D. Ramsay (for the Qweak Collaboration)}{
address={Department of Physics and Astronomy, University of Manitoba}
}
\begin{abstract} 

From a distance, the proton's weak charge is seen through the distorting
effects of clouds of virtual particles. The effective weak charge can be
calculated by starting with the measured weak coupling at the Z$^0$ pole and
``running'' the coupling to lower energy or, equivalently, longer distances.
Because the ``electroweak radiative corrections'' or ``loop diagrams'' which
give rise to the running depend not only on known particles, but on particles
which have not yet been discovered, a difference between the calculated and
measured weak charges may signal new physics. A measurement of $Q_{weak}$ to
4\% will be sensitive to new physics at the few TeV scale. The $Q_{weak}$
experiment is based on the fact that the parity-violating longitudinal
analyzing power, $A_z$, in electron-proton scattering at low momentum transfer
and small scattering angle, is proportional to the proton's weak charge. The
experiment plans to measure the predicted $A_z$ of -0.3 ppm with a combined
statistical and systematic uncertainty of 2.2\%, corresponding to a total
uncertainty of 4\% in $Q_{weak}$. This requires a statistical precision of $\pm
5 \times 10^{-9}$, which can be achieved in 2200 hours with an 85\% polarized,
180 $\mu$A electron beam incident on a 0.35 m liquid hydrogen target. A
synchronous data acquisition system will integrate the detector current signals
over each spin state and extract the helicity correlated, parity violating
component.

\end{abstract}

\maketitle

\section{Introduction}

The $Q_{weak}$ experiment [1] (Fig. 1) plans to make a precision measurement of
parity violation in the elastic scattering of longitudinally polarized electron
from protons. A longitudinally polarized electron beam is passed through a 35
cm thick liquid hydrogen target and forward scattered electrons are detected
using a magnetic spectrometer and detector system. The experiment measures the
fractional difference in cross section for right-handed and left-handed
electron helicities. It is expected that $A_z = (\sigma^+ - \sigma^-)/(\sigma^+
+ \sigma^-) \approx -0.3$ ppm, the negative sign indicating that the cross
section is slightly higher for the left-handed helicity.

In conventional, parity conserving, electron scattering experiments the
effective probe is the photon, which couples to the ``normal'' electromagnetic
charge and current. Many such experiments have been done and the distribution
of electric charge and magnetism in the proton is quite well known. In parity
violating electron scattering experiments, on the other hand, the effective
probe is the Z-boson, which couples to the weak charge. The weak charge of the
proton has not been measured yet. To do this is the goal of the $Q_{weak}$
experiment.

It is well established that observed charges vary with the distance at which
the charge is measured. For example the electric charge on the electron is
given by the Particle Data Group as $1.60217653(14) \times 10^{-19}$ C, where
the (14) is the uncertainty in the last two digits -- a value of impressive
accuracy. The electromagnetic coupling, which is the square of this charge
expressed in dimensionless units, is $\alpha_{QED} = e^2/4 \pi \epsilon_0 \hbar
c = 1/137.0359911(46)$. This is indeed the value measured at a large distance,
corresponding to measurements made at very low momentum transfer (low Q). As
the momentum transfer is increased, corresponding to probing closer and closer
to the bare charge, the observed charge increases [2]. At $Q^2 = m_W^2$,
corresponding to the mass of the W-boson, $\alpha_{QED} \sim 1/128$. This
dependence on distance is referred to as ``running''. The physical reason for
the running is that the bare charge is seen through the distorting effect of
clouds of virtual particles. In the case of QED, fermion pairs ``screen'' the
bare charge and cause it to appear smaller at larger distances. In the case of
the strong coupling of the non-abelian QCD, the behavior is the opposite. The
strong coupling is observed to be larger at longer distances and very weak
close up. The screening behavior in QED and the anti-screening in QCD have been
both calculated and confirmed by experiment [2,3]. The situation for the weak
charge is not as clear.

\begin{figure}
\includegraphics[width=120mm]{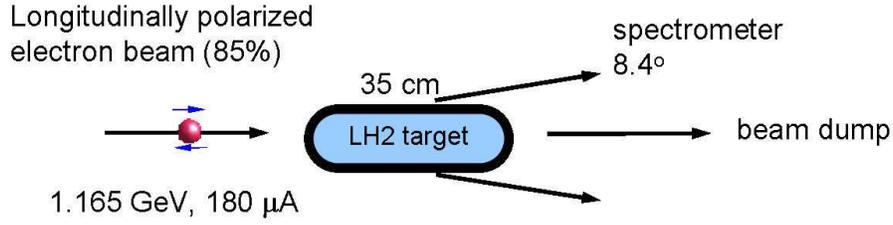}
\caption{Principle of the $Q_{weak}$ experiment. The elastic scattering cross
section for electrons on protons is slightly higher if the incident electron is
spinning to the left (negative helicity). At the forward angles and low
momentum transfer of the $Q_{weak}$ experiment, this difference is proportional
to the weak charge of the proton.}
\label{principle}
\end{figure}

The weak charge of the proton depends on the weak mixing angle, given by
$\theta_W = \tan^{-1} (g^{\prime} / g)$, where $g{\prime}$ is the U(1) gauge
group coupling and $g$ the SU(2) gauge group coupling, or by $\cos \theta_W =
M_W/M_Z$, with $M_Z$ and $M_W$ the masses of the Z and W bosons. The actual
numerical value of $\sin ^2 \theta _W$ depends on the renormalization scheme.
In terms of the weak mixing angle, the proton's weak charge is
$Q^p_{weak}=1-4\sin ^2 \theta _W$ + corrections. The corrections depend on how
much has been included in the definition of $sin ^2 \theta _W$ (i.e. on the
renormalization scheme). Figure 2 shows the running of $sin ^2 \theta _W$
calculated in the $\overline{MS}$ (``MS bar'') renormalization scheme [4]. Note
that large values of $sin ^2 \theta _W$ correspond to small values of
$Q^p_{weak}$. Also shown on the figure are some existing measurements [5] and
the uncertainty of the proposed $Q^p_{weak}$ measurement.

Because the radiative corrections, or loop diagrams, which give rise to the
running, depend not only on known particles, but on particles which have not
yet been discovered, a departure from the theoretical predictions could
indicate new physics. Our proposed measurement of $Q^p_{weak}$ to 4\%
corresponds to 0.3\% in $sin ^2 \theta _W$ and would be sensitive to new
physics on the few TeV scale. On the other hand, agreement with the
calculations would put strong constraints on Standard Model extensions.

\begin{figure}[ht]
\includegraphics[width=140mm]{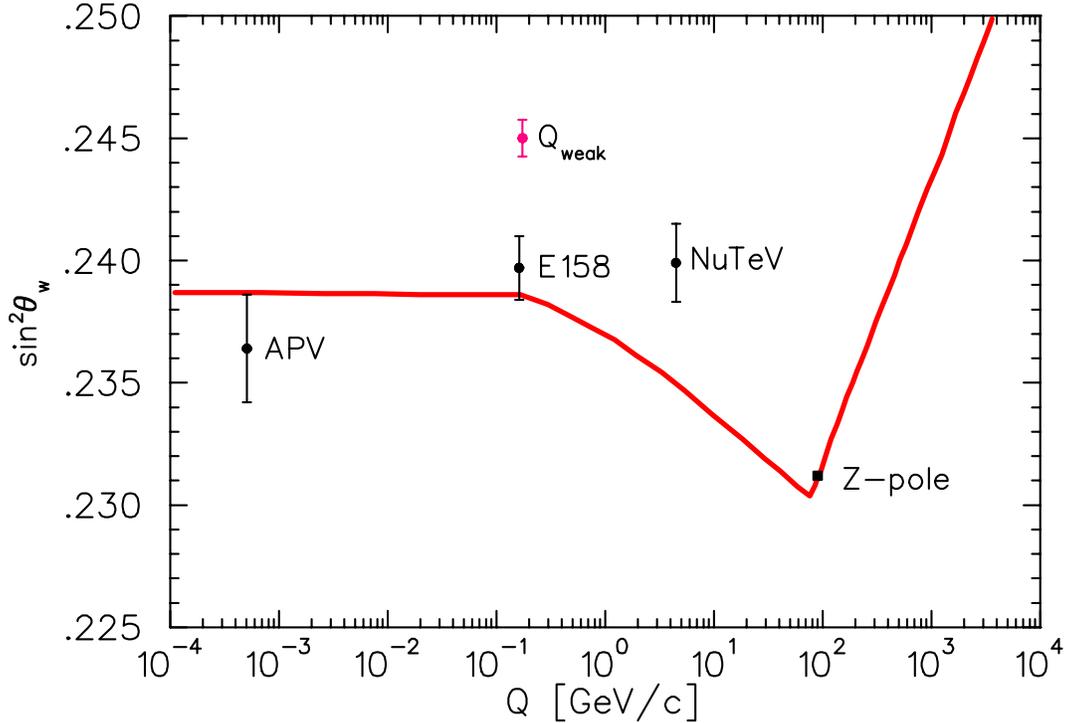}
\caption{ Running of $sin ^2 \theta _W$ calculated in the $\overline{MS}$
scheme. [4] Large $sin ^2 \theta _W$ corresponds to small $Q^p_{weak}$. The
lowest point of the curve is at the W-boson mass. At lower energy (longer
distance) we see screening of the weak charge, and at higher energy (shorter
distance) we see anti-screening. Existing measurements are shown in black, with
published statistical and systematic errors added in quadrature, and the
proposed $Q^p_{weak}$ measurement is shown in red with its expected
uncertainty. The vertical position of $Q^p_{weak}$ is arbitrary.}
\label{running}
\end{figure}

\section{The Experiment}

\subsection{Principle}

At forward angles and low Q$^2$, the parity violating analyzing power is [6]

\begin{displaymath}
A_z = \frac{-G_F}{4 \pi \alpha \sqrt{2}}{(Q^2Q^p_{weak} + Q^4 B)}.
\end{displaymath}

where $G_F$ is the Fermi coupling and $\alpha$ the fine structure constant.
Putting in the numbers and expressing the analyzing power in parts per million
(ppm) and the momentum transfer, Q, in GeV/c, gives

\begin{displaymath}
A_z = -90(Q^2Q^p_{weak} + Q^4 B).
\end{displaymath}

The first term, proportional to Q$^2$, is for a point-like proton. The second
term, proportional to Q$^4$, is a correction involving hadronic form factors.
Ideally we would like to run at low enough Q$^2$ that the proton would look
like a point and hadronic corrections would be negligible. Unfortunately, lower
Q$^2$ also reduces our signal, so some compromise is needed. We will use Q$^2$
= 0.03 (GeV/c)$^2$ and a scattering angle of 8 degrees. Based on standard model
calculations and global fits to existing hadronic data [7], we expect

\begin{displaymath}
A_z = -0.194 ppm -0.074 ppm = -0.268 ppm .
\end{displaymath}

The second term will be constrained by results from JLab (Gzero, HAPPEX), Mainz
(PV-A4), and MIT-Bates (SAMPLE), so by measuring $A_z$, we can extract the weak
charge, $Q^p_{weak}$.  

\subsection{Equipment}

Figure 3 shows the main parts of the Qweak experiment. The 1.165 GeV electron
beam, longitudinally polarized to more than 85{\%}, enters from the left and
passes through a 35 cm long liquid hydrogen target. Electrons scattered at 8
degrees pass through a series of collimators and an 8-sector toroidal magnetic
spectrometer to the main detectors. These are eight bars of synthetic quartz
each fitted at both ends with photomultipliers. Quartz was chosen because it is
radiation hard (we expect >300 krad) and is insensitive to gamma, neutron and
pion backgrounds. The bars should operate essentially at counting statistics.
The main detector region will be enclosed in a shielding house. This has been
removed in the figure to show the detectors. The Luminosity monitors are
located at very small forward angles where the analyzing power is almost zero.
They will monitor variations in beam current and also look for effects of
target boiling.

\begin{figure}
\includegraphics[width=140mm]{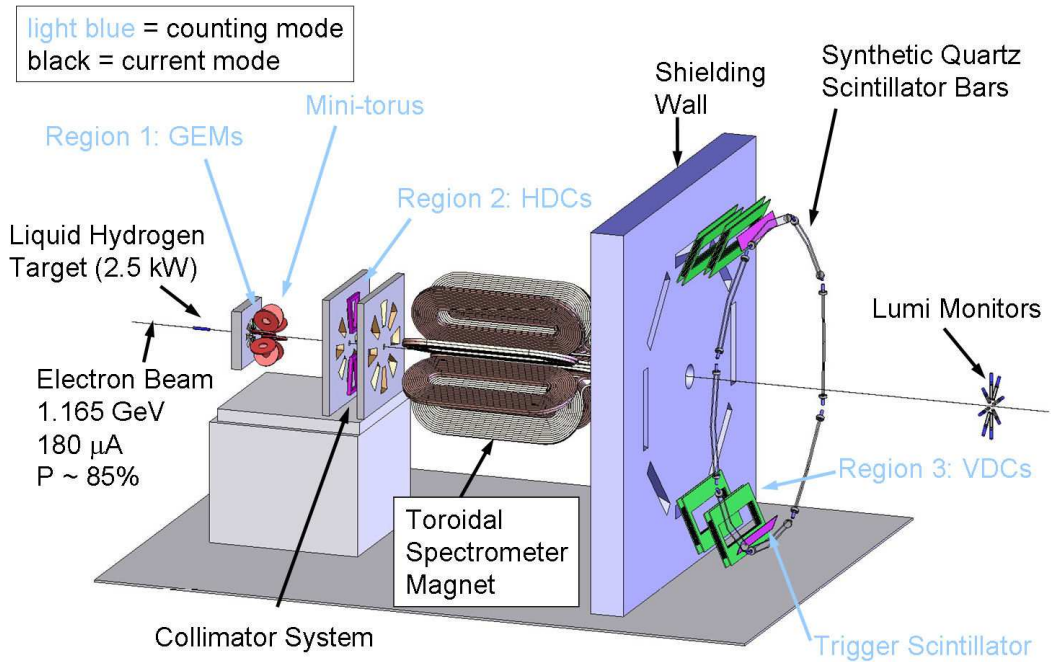}
\caption{Major parts of the Qweak experiment. For the main data taking runs the
position sensitive detectors in regions 1, 2, and 3 are not used. They slide
into place only for counting mode calibration runs, used to determine the
absolute $Q^2$. Pairs of chambers can be rotated to check two opposing main
detectors at a time, thus four runs are required to calibrate the whole array.}
\label{parts}
\end{figure}

The spectrometer is arranged such that inelastic electrons are deflected out
away from the main detectors and positive pions are defected inward. The
collimator is designed so that neutral particles hit the shield house inside
the detector window.

The liquid hydrogen target will be the highest power cryotarget ever. It will
be 35 cm long with a beam heat load of 2200 watts and a total heat load of 2500
watts. To achieve the cooling, plans are to use a 500 W auxiliary heat
exchanger using the end station refrigerator and a 2000 W heat exchanger using
the JLab central helium liquefier extra capacity.

Also shown in Fig. 3 in the locations marked Region 1, Region 2 and Region 3,
are position sensitive detectors for dedicated low current ($\sim$10 nA)
counting-mode calibration runs which will be occasionally made to determine the
absolute $Q^2$ and to study the backgrounds. Since the calibration is a
secondary measurement, only two chambers are used at each location. The pairs
can be rotated to measure all eight octants in 4 runs. When not in use, the
chambers will be retracted.

The main technical issues to be addressed can be seen from the expression for
$A_z$:

\begin{equation}
A_z = \frac{1}{P_z}\left(\frac{N^+ - N^-}{N^+ + N^-}\right)
=-90(Q^2Q^p_{weak}+Q^4B)
\end{equation}

\subsubsection{Statistics: N$^+$ and N$^-$}

\begin{figure}
\includegraphics[width=140mm]{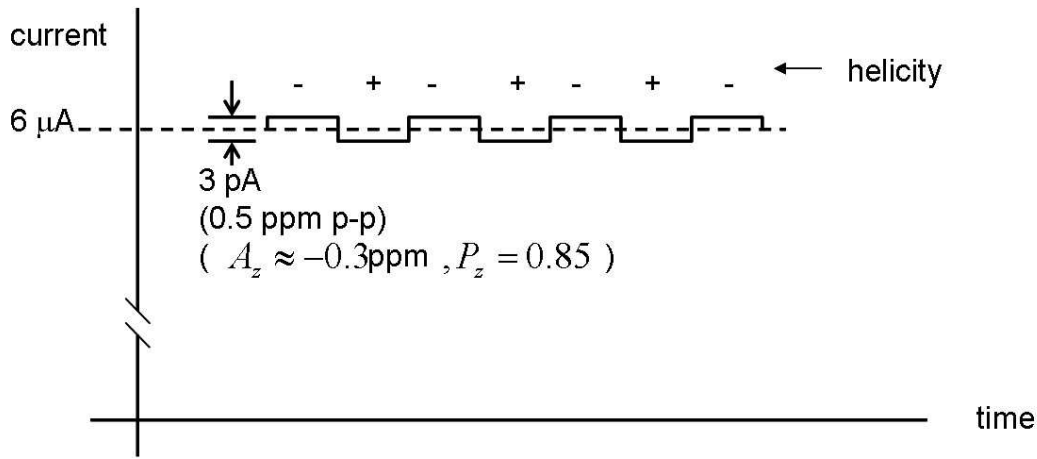}
\caption{Small size of the parity violating signal. The 6 $\mu$A signal from
the main detector is expected to be $\sim$3 pA greater in the negative helicity
state than in the positive helicity state (assuming a parity violating
analyzing power of -0.3 ppm and a longitudinal beam polarization of 85\%). Note
the suppressed origin. If you are reading this article on letter-sized paper,
the origin is 6 km off the bottom of the page.}
\label{signal}
\end{figure}

Our goal is to measure $A_z$ with a combined statistical and systematic
uncertainty of 2\%, corresponding to $\approx 4$\% on $Q^p_{weak}$, or $\approx
0.3$\% on $\sin ^2 \theta _W$. We are planning for $5 \times 10^{-9}$
statistics. To achieve this in our expected 2200 hours of running demands a
count rate of 6.4 GHz, too high for conventional particle counting techniques.
For this reason the main data taking for the $Q^p_{weak}$ experiment will be
done in current mode, using eight detectors running at 800 MHz each. The
detectors are 200 cm x 18 cm x 1.25 cm synthetic quartz bars, each with two
gain 1000 photomultipliers (PMTs), one at each end of the bar. Simulations and
preliminary tests indicate a signal of approximately 6 $\mu$A from each PMT.
Superimposed on this will be a very small parity violating signal synchronized
with the spin state. The small size of this signal is illustrated in Fig. 4.
This signal will be integrated over each spin state by precision digital
integrators being designed and built at TRIUMF. A synchronous data acquisition
system will then extract only the helicity-correlated part.

\subsubsection{Beam Polarization: $P_z$}

We expect a longitudinal beam polarization of >85\%. As seen from equation (1),
any uncertainty in the beam polarization, $P_z$, will appear directly as the
same fractional uncertainty in $A_z$. In order to determine the beam
polarization with an absolute uncertainty of less that 1\%, Jefferson Lab is
installing a new Hall-C Compton polarimeter. This polarimeter will be able to
run continuously during data taking, unlike the existing Møller polarimeter,
for which dedicated runs must be made. The Compton polarimeter should have no
difficulty achieving a statistical precision of better than 1\%, but it will
have to be cross-calibrated against the Møller to get the absolute accuracy.

\subsubsection{Absolute $Q^2$}

Since our desired $Q^p_{weak}$ appears multiplied by the square of the momentum
transfer, any error in $Q^2$ is a corresponding error in our extracted
$Q^p_{weak}$. For this reason the experiment is being built with the capability
to run in particle-counting, full-tracking mode. By reducing the beam current
to 10 nA and performing dedicated runs in counting mode with full tracking, we
will be able to make an absolute determination of $Q^2$ to 0.5\%. We will also
require a detailed field map of the spectrometer magnet. Dedicated runs
including time-of-flight will be used to estimate our background fraction.

\subsubsection{Hadronic Correction: $Q^4B$}

The $Q^4$ term will be estimated from a combination of theory and existing
experiments. A considerable body of data already exists on the hadronic form
factors, and by the time Qweak is running, more results will be available from
the JLab G zero experiment. Our current estimates are that hadronic structure
uncertainties will contribute 1.9\% uncertainty to our $Q^p_{weak}$ value.

\subsubsection{Helicity Correlated Beam Properties}

Our signal is a change in the detector current on helicity flip. If the only
thing that changed on helicity flip were the helicity, then the change in
detector current would be the true parity violating signal. Unfortunately, the
detector signal also depends on beam parameters other than helicity. For
example, it obviously depends directly on beam current; it will also depend to
a lesser degree on parameters such as beam position, beam size, and beam
energy. Changes in such beam properties, when coherent with spin flip, can
imitate parity violation and introduce a systematic error in our measurement.
The approach to minimizing such systematic errors is threefold. First, the
experiment is designed in such a way as to minimize the sensitivity to helicity
correlated beam properties. Second, machine conditions are carefully optimized
to minimize unwanted changes on helicity flip. In some cases active feedback
may be useful. Third, the actual helicity correlated beam properties are
measured during running, the sesitivities to these properties are measured, and
corrections are made for the resultant systematic error. Of course, each
systematic error correction carries with it an uncertainty which must be taken
into account when quoting the final result. We have carried out Monte Carlo
simulations to estimate our sensitivities to coherent modulations and have set
acceptable limits on the beam parameters. Table 1 shows the results of GEANT
simulations.

\begin{table}[!h]
\begin{tabular}{lccc} 
\hline\hline
  \tablehead{1}{c}{b}{Source} &
  \tablehead{1}{c}{b}{Error goes as} &
  \tablehead{1}{c}{b}{DC Conditions} &
  \tablehead{1}{c}{b}{Helicity Correlated}\\
\hline
Poition Modulation & $x_0\:\delta x$ & $x_0 < 0.7$\ mm   & $\delta x$\ = 20 nm \\
                   &               & $\delta r < 19$\ mm &                   \\
                   &               & $\delta B/B < 1.6$ \% &                 \\
Size Modulations   & $D_0\:\delta D$ & $D_0=200\ \mu$m & $\delta D<0.7\ \mu$m  \\
Direction Modulation & $\theta_0\:\delta\theta$ & $\theta_0 = 60\ \mu$rad &
$\delta\theta < 1.4\ \mu$rad \\
Energy Modulation & $\delta E$ & $E=1.165$ GeV & $\delta E = 7$ eV \\
 
\hline\hline
\end{tabular}
\caption{Helicity Correlated Beam Property constraints to keep false $A_z < 6
\times 10^{-9}$ based on GEANT simulations.}
\label{beamprop}
\end{table}

The conditions shown are those required to keep each false Az contribution to <
$< 6 \times 10^{-9}$. Notice that conditions can be traded off. For example if
the beam can be held closer to the neutral axis, then more beam motion can be
tolerated. The sensitivity to position modulation also depends on the symmetry
of the apparatus. As noted in column three, this simulation assumed that the
detector bars are positioned radially to better than 19 mm and that the
magnetic field is known to 1.6 {\%}. We do not anticipate any serious problem
meeting the beam quality specifications as the G zero experiment has already
achieved a similar quality of beam in Hall-C [8].

\subsubsection{Error Budget}

\begin{table}[!h]
\begin{tabular}{lcc} 
\hline\hline
  \tablehead{1}{c}{b}{Source} &
  \tablehead{1}{c}{b}{$\Delta A_z/A_z$} &
  \tablehead{1}{c}{b}{$\Delta Q_w/Q_w$}\\
\hline
Statistical (2200 hours)            &   1.8\%  &   2.9\%  \\
Systematic:                         &          &          \\
Hadronic Structure Uncertainties    &    --    &   1.9\%  \\
Beam Polarization                   &   1.0\%  &   1.6\%  \\
Absolute $Q^2$ determination        &   0.5\%  &   1.1\%  \\
Backgrounds                         &   0.5\%  &   0.8\%  \\
Helicity correlated beam properties &   0.5\%  &   0.8\%  \\
{\bf Total:}                        &   2.2\%  &   4.1\%  \\
 
\hline\hline
\end{tabular}
\caption{Qweak Error Budget. 2\% of $A_z \approx 4$\% of $Q_w \approx 0.3$\% of
$\sin ^2 \theta _W$}
\label{errors}
\end{table}

The expected contributions of various sources of uncertainty are summarized in
Table 2. The errors shown will lead to a 0.3\% determination of $\sin ^2 \theta
_W$. Actually, the raw uncertainty in $\sin ^2 \theta _W$ is closer to 0.2\%,
but an additional uncertainty associated with QCD corrections applied to the
extraction of $\sin ^2 \theta _W$ raises the uncertainty to 0.3\%.

\subsection{Status of the Experiment}

The Qweak collaboration was formed in May 2000. The JLab proposal was
approved with an ``A'' rating in January 2002 and the Technical Design
Review was completed in January 2003. In 2003 and 2004 funding was approved
by DOE, NSF and NSERC. In January 2005 a further JLab ``Jeopardy'' proposal
was approved, again with ``A'' rating. Here is a summary of the state of the
major sub-systems:

\begin{itemize}

\item All the magnet parts are at MIT and have been assembled and surveyed. We
expect to power up the magnet in the summer of 2007 and perform a magnetic
field map. Once this is complete the magnet can be delivered to Jefferson Lab.
The magnet should ship to JLab in the summer of 2008.

\item The first prototype digital integrator for the main current-mode running
has been tested at TRIUMF and shipped to JLab for further testing. Following
tests in 2007, more digital integrators and low noise preamplifiers will be
built at TRIUMF and delivered to JLab.

\item All the quartz bars needed for the 8 main detectors are now at JLab and
are undergoing quality control testing. Work is also proceeding at JLab on
design and testing of the low-gain photomultiplier and base package.

\item Design is proceeding well on the liquid hydrogen target. Work now is
concentrating of heat exchanger design.

\item Prototypes of most of the tracking chambers have been built and are being
tested.

\item A luminosity monitor (lumi) will be tested at JLab in 2007.

\item JLab engineers have produced a full 3D CAD model of the experiment. This
will be vital to verify the interfaces between different parts of the
experiment and the fitting of the experiment in Hall-C.

\end{itemize}

Installation of the experiment in Hall-C is scheduled to begin in March,
2009.

\begin{theacknowledgments}

This work is supported in part by the US DOE, NSF, NSERC (Canada), Jefferson
Laboratory and TRIUMF.

\end{theacknowledgments}


\begin{thebibliography}{}

\bibitem{Carli05} The Qweak Collaboration: R. Carlini et al., Jefferson Lab
Proposal E05-008 (2005).

\bibitem{Livi97} for example: I. Levine et al., Phys. Rev. Lett. 78, 424,
(1997).

\bibitem{Part06} see Particle Data Group review: J. Phys. G 33, 116, (2006).

\bibitem{Erle05} J. Erler and M.J. Ramsey-Musolf, Phys Rev. D 72, 073003
(2005).

\bibitem{Cesium} Atomic Cesium: S.C. Bennett and C.E. Wieman, Phys. Rev. Lett.
82, 2484, (1999); C.S. Wood et al., Science 275, 1759 (1997).\\
SLAC E158: P.L. Anthony et al., Phys. Rev. Lett. 95, 081601 (2005).\\
NuTeV: G.P. Zeller et al., Phys. Rev. Lett. 88, 091802 (2002).

\bibitem{Muso94} M.J. Musolf {\em et al.}, Physics Reports {\bf 239}, 1 (1994).

\bibitem{Youn06} R.D. Young, Jefferson Lab., private communication, (2006).

\bibitem{Naka05} K. Nakahara, Eur. Phys. J. A {\bf 24}, 119 (2005).

\end{thebibliography}
\end{document}